\documentclass{arima-en}
\pdfoutput=1 

\usepackage{array}
\usepackage{pgfplots}
\usepackage{graphicx}
\usepackage[utf8]{inputenc}
\usepackage[T1]{fontenc}
\usepackage{fancyhdr}
\usepackage{hyperref}
\usepackage[babel]{csquotes}

\usepackage[utf8]{inputenc} 
\usepackage[english]{babel}  
\usepackage{adjustbox}
\usepackage{float}
\usepackage{amsmath,amssymb,latexsym,amsfonts}
\usepackage{relsize}
\usepackage{flushend}
\usepackage{balance}
\usepackage{url}
\usepackage{multirow}
\usepackage[plain]{algorithm}
\usepackage{algorithmic} 
\usepackage{pgfplots}
\usepackage[babel]{csquotes}
\usepackage{setspace}   
\usepackage[font={small,stretch=0.84},
            labelfont={bf}
            ]{caption}  

\usepackage{setspace}
\usepackage{fancyhdr}
\usepackage{graphicx} 
\usepackage{fancybox,exscale,rotate,epsfig}
\usepackage{pifont}
\usepackage{rotating}
\usepackage{setspace}
\newcommand{\indic}{{\mathrm\mathbf1}}
\newcommand{\R}{\mathbb{R}}
\newcommand{\eqdef}     {\stackrel{\textup{\tiny def}}{=}}
\newcommand{\rmd}{{\textup{d}}}

\newcommand{\E}{\mathbb{E}}
\newcommand{\FF}{\mathcal{F}}
\newcommand{\NN}{\mathcal{N}}
\newcommand{\MM}{\mathcal{M}}
\makeatletter
\newcommand{\oset}[3][0.1ex]{%
  \mathrel{\mathop{#3}\limits^{
    \vbox to#1{\kern-2.3\ex@
    \hbox{$\scriptstyle#2$}\vss}}}}
\makeatother
\newcommand{\simiid}     {\oset{\textrm{\tiny iid}}{\sim}}

\usepackage{algorithm,algorithmic,amssymb,amsmath,letltxmacro}
\newlength{\commentindent}
\makeatletter
\renewcommand{\algorithmiccomment}[1]{\unskip\hfill\makebox[\commentindent][l]{\tt\#~#1}\par}
\LetLtxMacro{\oldalgorithmic}{\algorithmic}
\renewcommand{\algorithmic}[1][0]{%
  \oldalgorithmic[#1]%
  \renewcommand{\ALC@com}[1]{%
    \ifnum\pdfstrcmp{##1}{default}=0\else\algorithmiccomment{##1}\fi}%
}
\makeatother

\usepackage[
  style=numeric-comp,
  datamodel=software, 
  abbreviate=false,
  natbib=true,
  backend=biber,
  bibencoding=utf8,
  giveninits=true,
  url=false,
  doi=false,
  defernumbers,
  maxcitenames=10,
  defernumbers=false,
  maxbibnames=100]{biblatex}
\usepackage{software-biblatex}
\title{The one step fixed-lag particle smoother \\
as a strategy to improve\\ 
the prediction step of particle filtering}

\author[*1,2]{Samuel Nyobe}
\author[3]{Fabien Campillo}
\author[1,2]{Serge Moto}
\author[4,5]{Vivien Rossi}

\affil[1]{MIBA research Unity, Faculty of Science,
University of Yaoundé, Cameroon}
\affil[2]{UMI 209, UMMISCO-Cameroon, IRD,
Sorbonne University,  France}
\affil[3]{Inria, MathNeuro Team, Montpellier, France}
\affil[4]{RU Forets and Societies, CIRAD, Yaoundé, Cameroon}
\affil[5]{National Advanced School of Engineering, University of Yaoundé, Cameroon}

\corrauthor{samuel.nyobe@facsciences-uy1.cm}

\doi{10.46298/arima.10784}{}
\review{6 janvier 2023}{30 novembre 2023}
\publication{39}{2023}
\editors{Bruce Watson, Mathieu Roche, Nabil Gmati, Clémentin Tayou Djamegni}
\usepackage{lastpage}
\usepackage{software-biblatex}
\newcommand{\doiorurl}{%
 \iffieldundef{doi}
    {\iffieldundef{url}
       {}
       {\strfield{url}}}
    {http://dx.doi.org/\strfield{doi}}%
}
\newcommand{\myhref}[1]{%
 \ifboolexpr{%
  test {\ifhyperref}
  and
  not test {\iftoggle{bbx:url}}
   and
   not test {\iftoggle{bbx:doi}}
  }
  {\href{\doiorurl}{#1}}
  {#1}%
}
\DeclareFieldFormat{title}{\myhref{\mkbibemph{#1}}}
\DeclareFieldFormat
  [article,inbook,incollection,inproceedings,patent,thesis,unpublished]
  {title}{\myhref{\mkbibquote{#1\isdot}}}
\pgfplotsset{compat=1.17}
\begin{document}
\maketitle 


\abstract{
Sequential Monte Carlo methods have been a major advance in the field of 
numerical filtering for stochastic dynamical state-space systems with 
partial and noisy observations. However, these methods  still have some weaknesses. 
One of its main weaknesses concerns the degeneracy of  these particle filters due to 
the impoverishment of the particles. 
Indeed, during the prediction step of these filters, the particles explore the state space, 
and if this exploration phase is not done correctly, a large part of the particles will end up 
in  areas that are weakly weighted by the new measurement 
and will be mostly eliminated. Only a few particles will remain, leading to a degeneracy of the filter.
In order to improve this last step within the 
framework of the classic bootstrap particle filter, we propose a simple
approximation of the one step fixed-lag smoother. At each time 
iteration, we 
propose to perform additional simulations during the prediction step in order to 
improve the likelihood of the selected particles. 
Note that we aim to propose an algorithm that is almost as fast and of the same order of 
complexity as the bootstrap particle filter, and which is robust in poorly conditioned filtering situations.
We also investigate a robust version of this smoother.
}


\section{Introduction}
\label{introduction}

Since the 1980s sequential Monte Carlo (SMC) methods, also called particle 
filter (PF) methods \cite{gordon1993a, doucet2001a, arulampalam2002a, 
doucet2011a,crisan2011a}, have met with vast success and broad usage in 
the context of nonlinear filtering for state-space models with 
partial and noisy 
observations, also known as hidden Markov models (HMM) for state-space
models. 
The success of these methods is due to their ability to take into 
account, in a numerically realistic and efficient way, the non-linearity of the 
dynamics and the non-Gaussianity of the underlying conditional distributions.

The exact (i.e., non approximated) dynamics of these HMMs takes the form of a 
sequential Bayes formula represented in 
Eqs.~\eqref{eq.filtre.pred}-\eqref{eq.filtre.cor} 
also called Bayesian filter. 
In the case of linear models with additive Gaussian noise, 
the Kalman filter gives an exact solution of this problem as well as an efficient algorithm to compute it.
There are a few cases where this filter can be computed explicitly 
in a finite-dimensional way, but in the majority of cases it is necessary to use numerical approximation methods.
Historically, the first method proposed is the extended Kalman filter (EKF) \cite{anderson1979a} and its variants, which consist in linearizing the model around the current estimate and applying the Kalman filter. 
But in the case of strongly nonlinear models, the EKF often diverges.
Since the 1980s, Monte Carlo methods have become very effective alternatives,
they are now recognized as a 
powerful tool in the estimation with Bayesian filters in nonlinear/non-Gaussian hidden 
Markov models. 

Among Monte Carlo methods, the PF techniques rely on an online importance sampling 
approximation of the sequential Bayes formula \eqref{eq.filtre.pred}-\eqref{eq.filtre.cor}. At each time 
iteration, the PF 
builds a set of particles i.e. an  independently and identically 
distributed sample from an approximation of the theoretical solution of the Bayesian 
filter.

A PF time iteration classically includes two steps: 
a prediction step consisting in exploring the state space by moving the particles according to the state equation, 
followed by a correction step consisting in first weighting the particles according to their correspondence with the new observation, i.e. according to their likelihood, then resampling these particles according to their weight.
These two steps can be understood respectively as the mutation and selection steps of genetic algorithms.

The first really efficient PF algorithm, namely the bootstrap particle filter (BPF) or 
sampling-importance-resampling filter proposed in 1993 \cite{gordon1993a}, follows exactly these two steps.

The resampling step is essential,
without it we notice after a few  time iterations an 
impoverishment of the particles:
very few particles, even one, will concentrate all the likelihood; the other particles then become useless, the filter then loses track of the current state.

This resampling step is therefore essential, but still it does not prevent another type of particle degeneracy 
 \cite{arulampalam2002a, 
seongkeun_park2009a, doucet2011a, junyi_zuo2013a,tiancheng_li2014a}.
The latter appears when the particles explore areas of the state space that are not in agreement with the new observation.
This is due to the fact that in its classical version, especially in the case of BPF, the prediction step propagates the particles in the state space without taking into account the next observation.
Thus, when we associate weights with particles using their value via the likelihood function, a large proportion of the particles may have almost zero weight, 
this phenomenon is aggravated by the machine epsilon. 
In the worst case, all the particles can have a zero weight,
the filter has then completely lost  track of the state evolution.

To overcome this problem it is relevant to take into account the observations in the prediction step consisting in exploring the state space. 
The auxiliary particle filter (APF)  \cite{pitt1999a,pitt2001a} is one of the first attempts to take into account the observation in the prediction step.
One can consult the review articles \cite{godsill2019a, godsill2001b} concerning the possibilities of improvement of particle filters.

Another possibility to overcome this difficulty, without using the observation following the current one, is to use robust versions of particle filters, such as the 
regime switching method \cite{ellaham2021}  or the averaging-model approach 
\cite{liu2019,urteaga2016}, which allow  model uncertainties to be taken 
into account.

\medskip

The present paper aims to propose a new PF algorithm, called predictive bootstrap 
particle smoother (PBPS), to further improve the prediction step but keeping the simplicity of the BPF. The principle of the PBPS is to take into account the current and the next observations for the prediction and correction steps. At each prediction step, 
we make additional explorations in order to better determine the likelihood according to the next observation.
 It can be seen as a simple approximation of the one step fixed-lag smoother.
The additional explorations contribute to the correction step in order to improve the likelihood of the selected particles with the current and next observation.   

We  propose an algorithm that is only a little more complex than the BPF, which is why we do not consider more complex algorithms, like the smoother with a deeper time-lag. We also want to propose an efficient algorithm  in case of unfavorable filtering situations such as when observability is poorly conditioned. 

In Section \ref{sec.nonlinear.filtering.smoothing} 
state-space models and the (exact) formulas for nonlinear filters and smoothers are presented.
In Section \ref{sec.particle.approximations}, we first present the classical bootstrap particle filter (BPF), then the predictive bootstrap particle smoother  (PBPS) and its ``robustfied'' version called regime switching predictive bootstrap particle smoother (RS-PBPS).

In Section \ref{sec.simulation} we present simulations in two test cases where observability is weakly conditioned. 
The first test case is a classical example in state space dimension 1 for which we compare the PBPS to the extended 
Kalman filter  (EKF) \cite{anderson1979a}, the unscented Kalman filter (UKF) \cite{vandermerwe2000a,julier2004a}, the  bootstrap particle filter (BPF)~\cite{gordon1993a}, and the the auxiliary particle filter (APF)  \cite{pitt2001a}.
The second  is more realistic,  
it is a classical problem of bearing-only 2D tracking with a state space of 
dimension 4.
In this case we first compare the PBPS to the BPF and the APF filters. In 
this example there are often model uncertainties, so it is relevant to 
use robust filters, and we compare the RS-PBPS to ``robustified'' versions of the particle filters.

\section{Nonlinear filtering and smoothing}
\label{sec.nonlinear.filtering.smoothing}

In order to simplify the presentation, we will make the abuse of notation, customary in this field, of representing probability distributions, e.g. $\mu(\rmd x)$, by densities, e.g. $\mu(x)\,\rmd x$; we also represent the Dirac measure $\delta_{x_{0}}(\rmd x)$ as $\delta_{x_{0}}(x)\,\rmd x$ where $\delta_{x_{0}}(x)$ is the Dirac function 
(with $\delta_{x_{0}}(x)=0$ if $x\neq x_{0}$, $\delta_{x_{0}}(x)=+\infty$ if $x= x_{0}$, and $\int \delta_{x_{0}}(x)\,\rmd x =1$).

\subsection{The state space model}
\label{sec.state.space.model}

We consider a Markovian state-space model with state process 
$(X_k)_{k\geq 0}$ taking values in $\R^n$ and observation process  
$(Y_k)_{k\geq 1}$ taking values in $\R^d$. We suppose that 
conditionally on $(X_k)_{k\geq 0}$, the observations $Y_k$ are 
independent.

The ingredients of the state-space model are:
\begin{align*}
  q_k(x|x')     
  & \eqdef p_{X_k|X_{k-1} = x'}(x)\,,
  & & \textrm{(state transition kernel)} 
  \\
  \psi_{k,y}(x) 
  & \eqdef p_{Y_k|X_k = x}(y)   \,, 
  & & \textrm{(local likelihood function)} 
  \\
  \rho_0(x)      
  & \eqdef p_{X_0}(x)\,,
  & & \textrm{(initial distribution)} 
\end{align*}
for any $x,x'\in\R^n$, $y\in\R^d$.

\medskip

These ingredients can be made more explicit in the case of the following state space model:
\begin{align}
\label{eq.state.space.X}
  X_{k} 
  &=
  f_{k-1}(X_{k-1})+g_{k-1}(X_{k-1})\,W_{k-1}\,,
  \\
\label{eq.state.space.Y}
  Y_{k} 
  &=
  h_{k}(X_{k})+V_{k}\,,
\end{align}
for $k\geq 1$, where $X_{k}$ (resp. $Y_{k}$, $W_{k}$, $V_{k}$) takes values in $\R^n$
(resp. $\R^d$, $\R^m$, $\R^d$), $f_{k}$ (resp. $g_{k}$, $h_{k}$) is a differentiable 
and at most linear growth, uniformly in $k$, from $\R^n$ to $\R^n$ (resp. $\R^d$, $\R^{n\times m}$, 
$\R^d$), $g_{k}$ being also bounded. Random sequences $W_{k}$ and $V_{k}$ are independent 
centered white Gaussian noises with respective variances $R^W_{k}$ and $R^V_k$; 
$W_{k}$, $V_{k}$, $X_{0}$ being independent. In this case:
\begin{multline*}
    q_k(x|x')
    =
    \frac
    {1}
    {\sqrt{(2\,\pi)^n\,\det R^W_{k-1} }}\times
\\
\times\exp\Bigl(
         -\frac12 
         \bigl[x-f_{k-1}(x')\bigr]^* \, 
         \bigl[g_{k}(x')\,R^W_{k-1}\,g_{k}(x')^*\bigr]^{-1}\, 
         \bigl[x-f_{k-1}(x')\bigr]
        \Bigr)    
\end{multline*}
and
\[
   \psi_{k,y}(x) 
   \propto 
   \exp\Bigl(-\frac12 \bigl[y-h_{k}(x)\bigr]^* \, 
                   [R^V_{k}]^{-1}\, \bigl[y-h_{k}(x)\bigr]\Bigr)
\]
($\psi_{k,y}$ has to be known up to a multiplicative constant).

\subsection{Nonlinear filtering}

Nonlinear filtering aims at determining the conditional distribution $\eta_k$ of the current  state $X_{k}$ given the past observations $Y_{1},\dots,Y_{k}$, namely:
\begin{align*}
	\eta_k(x) \eqdef  p_{X_{k}|Y_{1:k} = y_{1:k}}(x) 
	\,,\ x\in\R^n
\end{align*}
for any $k\geq 1$ and $y_{1:k}\in(\R^d)^k$, here we use the  notation:
\[
 \textrm{``$Y_{1:k}$''  for $(Y_{1},\dots,Y_{k})$}
\]
(e.g. $Y_{1:k} = y_{1:k}$ means $Y_{\ell} = y_{\ell}$ for 
all $\ell=1,\dots,k$).

\medskip

The nonlinear filter allows us to determine $\eta_k$ from $\eta_{k-1}$ using the 
classical two-step recursive Bayes formula:
\begin{description}

\item[]
\quad\textbf{Prediction step.}
The predicted distribution $\eta_{k^{-}}(x) \eqdef  p_{X_{k}|Y_{1:k-1} = y_{1:k-1}}(x)$ of $X_{k}$ given $Y_{1:k-1}=y_{1:k-1}$ is given by:
\begin{align}
\label{eq.filtre.pred} 
	\eta_{k^{-}}(x) 
	= 
	\int_{\mathbb{R}^{n}}\ q_{k}(x|x')\ \eta_{k-1}(x')\ \rmd x'\,,
\end{align}
for $x\in\R^n$.

\item[]
\quad\textbf{Correction step.}
The new observation $Y_{k}=y_{k}$ allows   the predicted distribution 
to be updated in order to obtain $\eta_{k}$ according to the Bayes formula:
\begin{align}
\label{eq.filtre.cor} 
	 \eta_k(x) 
	 &= 
	 \frac
	 {\psi_{k}(x) \, \eta_{k^{-}}(x)}
	 {\int_{\R^n} \psi_{k}(x')\, \eta_{k^-}(x')\,\rmd x'}
	 \,,
\end{align}
for $x\in\R^n$; here and in order to simplify the notations, $\psi_{k}(x)$ represents $\psi_{k,y_{k}}(x)$.

\end{description}
\medskip

Note that in the correction step \eqref{eq.filtre.cor} , $\eta_k$ is proportional to 
the product of $\eta_{k-1}$ and the local likelihood function, that is: 
$$\eta_k(x) \propto  
\psi_{k}(x) \,\eta_{k^{-}}(x)\,.
$$

\subsection{Nonlinear smoothing}
\label{sec.smooth}

Define 
$\bar\eta_{k}(x)$ the conditional distribution of $X_{k}$ given $Y_{1:k+1}=y_{1:k+1}$, that is:
\begin{align*}
	\bar\eta_k(x) \eqdef  p_{X_{k}|Y_{1:k+1} = y_{1:k+1}}(x) 
	\,,\ x\in\R^n
\end{align*}
for any $k\geq 1$ and $y_{1:k+1}\in(\R^d)^{k+1}$.

\medskip

To get the recurrence equation for $\bar\eta_{k}(x)$, 
we consider the extended state vector  $\mathbf{X}_{k}=(X_{k},X_{k-1})$, it's a Markov with transition:
\begin{align*}
  &Q_{k}(x',x''|x_{k-1},x_{k-2})=
\\
  &\qquad=
  p_{X_{k},X_{k-1}|X_{k-1}=x_{k-1},X_{k-2}=x_{k-2}}(x',x'')
\\
  &\qquad=
  p_{X_{k}|X_{k-1}=x'',X_{k-1}=x_{k-1},X_{k-2}=x_{k-2}}(x')\;
  p_{X_{k-1}|X_{k-1}=x_{k-1},X_{k-2}=x_{k-2}}(x'')
\\
  &\qquad=
  p_{X_{k}|X_{k-1}=x_{k-1},X_{k-2}=x_{k-2}}(x')\;
  \delta_{x_{k-1}}(x'')
\\
  &\qquad=
  p_{X_{k}|X_{k-1}=x_{k-1}}(x')\;
  \delta_{x_{k-1}}(x'')
\\
  &\qquad=
  q_{k}(x'|x_{k-1})\;
  \delta_{x_{k-1}}(x'')\,,
\end{align*}
where $x''\to\delta_{x_{k-1}}(x'')$ is the Dirac delta function in $x_{k-1}$
(here, we adopt the usual abuse of notation, which consists in representing the Dirac measure by the Dirac delta function:
$\delta_{x}(x')=0$ if $x'\neq x$, $\delta_{x}(x')=+\infty$ if $x'= x$, and $\int \delta_{x}(x')\,\rmd x' =1$.)

\medskip

The conditional distribution $\boldsymbol{\eta}_{k}(x',x'')$ of $\mathbf{X}_{k}=(X_{k},X_{k-1})$ given $Y_{1:k}=y_{1:k}$, 
we apply the previous filter formula:
\begin{align*}
\boldsymbol{\eta}_{k^-}(x',x'')
 &=
  \iint Q_k(x',x''|x_{k-1},x_{k-2})\,\boldsymbol{\eta}_{k-1}(x_{k-1},x_{k-2})\,
\rmd x_{k-1}\,\rmd x_{k-2}
\\
  &=
  \iint q_k(x'|x_{k-1})\,\delta_{x_{k-1}}(x'')\,\boldsymbol{\eta}_{k-1}(x_{k-1},x_{k-2})\,
  \rmd x_{k-1}\,\rmd x_{k-2}
\\
  &=
  \iint q_k(x'|x_{k-1})\,\delta_{x_{k-1}}(x'')\,\boldsymbol{\eta}_{k-1}(x'',x_{k-2})\,
  \rmd x_{k-1}\,\rmd x_{k-2}\,,
\end{align*}
and the distribution of $Y_{k}$ given $(X_{k}=x_{k},X_{k-1}=x_{k-1})$ is the distribution of $Y_{k}$ 
given $X_{k}=x_{k}$, hence:
\begin{align}
\label{eq.smooth.corr}
  \boldsymbol{\eta}_{k}(x',x'')
  &\propto
  \psi_{k}(x')\,\boldsymbol{\eta}_{k^-}(x',x'')\,,
\end{align}

and the $x''$-marginal distribution of this last expression gives $\bar\eta_{k-1}(x'')$, the conditional distribution  of $X_{k-1}$ 
given $Y_{1:k}=y_{1:k}$.

\subsection{Approximations}

The main difficulty encountered by the nonlinear filter 
\eqref{eq.filtre.pred}-\eqref{eq.filtre.cor} lies in the two 
integrations. These integrations can be solved explicitly 
only in the linear/Gaussian case, leading to the Kalman filter,  
and in a very few other specific 
nonlinear/non-Gaussian cases. 
In the latter cases, the optimal filter can be solved \emph{explicitly} 
in the form of a finite dimensional filter; hence in the vast 
majority of cases, it is necessary to use approximation 
techniques~\cite{crisan2011a}. Among the approximation techniques, 
we will consider the extended Kalman filter (EKF) 
\cite{anderson1979a}, the  unscented Kalman filter (UKF) 
\cite{vandermerwe2000a,julier2004a} and  particle filter techniques, 
also called sequential Monte Carlo techniques 
\cite{gordon1993a,doucet2001a}, see \cite{doucet2018a} for a recent overview.

\medskip

Concerning the particle filters, as we will see in the next section, it is 
important to notice that on the one hand we do not need to know the 
analytical expressions of the state transition kernel and of the  
initial distribution, we just need to be able to sample (efficiently) 
from them; on the other hand, we do need the analytical expression of 
the local likelihood function (up to a multiplicative constant), indeed, 
for a given $y$, we need to compute $\psi_{k,y}(x)$ for a very large number 
of $x$ values.

\section{Particle approximations}
\label{sec.particle.approximations}

Particle filtering, also known as sequential Monte Carlo (SMC) methods, is a set of computational techniques used for the approximation of nonlinear filters \cite{doucet2001a}. According to this approach, the nonlinear filter $\eta_{k}$ is approximated by   $\eta_{k}^N$   of the form:
\begin{align*}
 \eta_{k}^N(x) &= \sum_{i=1}^{N}\omega^i_{k}\,\delta_{\xi_{k}^i}(x)
 \,,\ x\in\R^n\,.
\end{align*}
composed of $N$ particles located in $\xi_{k}^i$ in $\R^n$ 
and associated weights $\omega^i_{k}$, the weights are 
positive and sum to one \cite{doucet2001a}. Ideally the particles are sampled from 
$\eta_{k}$ 
and the importance weight are all equal to $1/N$ (meaning that all the particles have the same ``importance''). 

\medskip

Particle filters offer two essential advantages in numerical approximation:
\begin{itemize}
\item
they are finite-dimensional filters in the sense that they are represented using a finite number of parameters, in this case the position of the particles (after selection), which can therefore be integrated into computer memory; 
\item
the representation of particle filters, as weighted sums of Dirac measurements, 
greatly facilitates the calculation of integrals according to these filters; moreover, many operations can be parallelized as a function of the particle index 
(or vectorized with interpreted languages like \verb+Python+).
\end{itemize}

\medskip

In the 3 particle filters presented, at each time increment between two 
observations, in a first so-called ``mutation'' step, 
the particles explore the state space. A weight is then associated with each of these particles according to a fitness function (here a local likelihood function), these weights will be noted $\omega^{1:N}_{k}$ in the 3 cases, then in a second ``selection'' step, particules are selected based on the fact that particles with a high weight will have a greater probability of reproduction than particles with low weights (at constant number of particles $N$), leading to a family of particles noted $\xi^{1:N}_{k}$ in the 3 cases.

\subsection{The bootstrap particle filter}

The bootstrap particle filter  (BPF) filter is a classical sequential importance 
resampling method: suppose we have a good approximation 
$\eta_{k-1}^N=\frac{1}{N}\sum_{i=1}^{N}\delta_{\xi_{k-1}^i}$ of $\eta_{k-1}$. We can 
apply the prediction step \eqref{eq.filtre.pred} to $\eta_{k-1}^N$ and get the
following approximation of $\eta_{k^-}$:
\begin{align*}
    \tilde\eta_{k^-}(x)
    \eqdef
    \int_{\mathbb{R}^{n}}\ q_{k}(x|x')\ \eta_{k-1}^N(x')\, \rmd x' 
	= \sum_{i=1}^{N}\omega_{k-1}^i\ q_k(x|\xi_{k-1}^i)\,,
\end{align*}
and then apply the correction step \eqref{eq.filtre.cor}  and get the
following approximation of $\eta_{k}$:
\begin{align*}
    \tilde\eta_{k}(x)
    \eqdef
   \frac
	{\sum_{i=1}^{N}
	  \omega_{k-1}^i \, \psi_{k}(x) \, q_k(x|\xi_{k-1}^i)}
	{\int_{\R^n}
	\sum_{i=1}^{N}
	  \omega_{k-1}^i \, \psi_{k}(x') \, q_k(x'|\xi_{k-1}^i)\,\rmd x'}
	\,, 
\end{align*}
but the two last approximations $\tilde\eta_{k^-}$ and $\tilde\eta_{k}$
are mixtures of the continuous densities 
$q_k(\,\cdot\,|\xi_{k-1}^i)$ 
and therefore not of the particle type. The bootstrap particle filter (BPF) 
proposed by \cite{gordon1993a} is the simplest method to propose a particle 
approximation. For the prediction step we use the sampling technique:
\begin{align*}
   \xi^i_{k^-} \sim q_k(\,\cdot\,|\xi^{i}_{k-1})\,,\ i=1:N\quad \,\textrm{(independently)}\,,
\end{align*}
which is:
\[
   \xi^i_{k^-}
   =
   f_{k-1}(\xi^{i}_{k-1})
   +
   g_{k-1}(\xi^{i}_{k-1})\,\textrm{w}^i\,,
\]
where $\textrm{w}^i$ are i.i.d. $N(0,R^W_{k-1})$ samples. Then we let
\begin{align*}
   \eta_{k^-}^N(x) 
   \eqdef 
   \sum_{i=1}^{N}\omega_{k-1}^i\,\delta_{\xi_{k^-}^i}(x)\,.
\end{align*}
Through the correction step \eqref{eq.filtre.cor}, this approximation  $\eta_{k^-}^N$ gives $\sum_{i=1}^{N}\omega_{k}^i\,\delta_{\xi_{k^-}^i}$ where:
\begin{align}
\label{eq.sir.likelihood}
  \omega^i_k \eqdef 
  \frac
  {\psi_k(\xi_{k^-}^i)\,\omega_{k^-}^i}
  {\sum_{j=1}^N\psi_{k}(\xi_{k^-}^j)\,\omega_{k^-}^j}
\end{align}
are the updated weights taking account of the new observation $y_{k}$ through the 
likelihood function $\psi_{k,y_k}$.
This correction step \textit{must} be completed by a resampling of the particles 
$\xi_{k^-}^i$ 
according to the importance weights $\omega^i_k$:
\begin{align}
\label{eq.sir.ressampling}
  \xi_{k}^i \sim \sum_{j=1}^{N}\omega_{k}^j\,\delta_{\xi_{k^-}^j}
  \,,\ i=1:N\quad \textrm{(independently)}\,,
\end{align}
leading the to the particle approximation:
\begin{align*}
   \eta_{k}^N(x) 
   \eqdef 
   \sum_{i=1}^{N}\omega_{k}^i\,\delta_{\xi_{k}^i}(x)\,,\ \textrm{with }
   \omega_{k}^i = \frac1N\,.
\end{align*}
Algorithm \ref{algo.bpf} gives a summary of the BPF, in this version a resampling is performed at each time interval. The multinomial resampling step \eqref{eq.sir.ressampling}, whose basic idea is proposed 
in \cite{gordon1993a}, could be greatly improved, there are several resampling techniques, see \cite{douc2005a,tiancheng_li2015a} for more details.

\begin{listing}
\small
\setstretch{1.3}
\begin{algorithmic}[1] 	    	    
\STATE $\xi^{1:N}_{0} \simiid \rho_{0}$ 
	\COMMENT{initialization}
\RETURN $\xi_0^{1:N}$
\FOR{$k = 1,2,\dots$}  
	\STATE $\xi^{i}_{k^-} \sim q_k(\,\cdot\,|\xi^{i}_{k-1}) \,,\  i = 1:N$ 
		\COMMENT{particles evolution}
	\STATE $\omega^i_{k} \leftarrow \psi_{k}(\xi^{i}_{k^-}) \,,\  i=1:N $ 
		\COMMENT{likelihood}
	\STATE $\omega^i_k \leftarrow  
		{\omega^i_k}/{\sum_{j=1}^{N}\omega^{j}_k} \,,\  i=1:N$ 
   		\COMMENT{renormalization}
	\STATE $\xi_k^{1:N} \simiid 
		\sum_{j=1}^{N}\omega^{j}_k\,\delta_{\xi^{j}_{k^-}}$
   		\COMMENT{resampling} 	
    \RETURN $\xi_k^{1:N}$	 
\ENDFOR
\end{algorithmic}
\caption{Bootstrap particle filter (BPF).}
\label{algo.bpf}
\end{listing}

\bigskip

It is not necessary to resample the particles  
 systematically at each time iteration as in Algorithm \ref{algo.bpf}. 
However, this resampling step should be done regularly in terms 
of time iterations to avoid degeneracy of the weights \cite{doucet2001a}.

We will consider another degeneracy problem. Suppose that in step 
\eqref{eq.sir.likelihood}, the local likelihood values 
$\psi_{k}(\xi_{k^-}^i)$ 
associated with the predicted particles $\xi_{k^-}^i$ are all very small, 
or even equal to zero due to rounding in floating point arithmetic. This 
normalization step \eqref{eq.sir.likelihood} is then impossible. This problem 
occurs when the filter loses track of the true state $X_k$, i.e. the 
likelihood of the predicted particles $\xi_{k^-}^i$ w.r.t. the observation 
$y_{k}$ are all negligible, in this case the observation $y_{k}$ appears as an 
outlier. This occurs especially when the period of time between two successive 
instants of observation is very large compared to the dynamics of the state 
process. Strategies to  overcome this weakness encompass 
the  iterated extended Kalman particle filter   
\cite{liang_qun2005a} or the iterated unscented Kalman particle filter  
\cite{guo2007a}; see \cite{godsill2001b} and \cite{tiancheng_li2014a} for 
reviews of the subject.

\subsection{The predictive bootstrap particle smoother}
\label{b2_filter}

The purpose of the predictive bootstrap filter (PBPS) is to improve the 
correction step \eqref{eq.sir.likelihood} of the BPF by using both observations 
$Y_k = y_k$ and  $Y_{k+1} = y_{k+1}$ at each iteration $k$. 
In a way, the PBPS can be seen as an approximation of the distribution 
of $X_{k}$ given $Y_{1:k+1} = y_{1:k+1}$, the one step fixed-lag smoother presented 
in Section \ref{sec.smooth}. 

\medskip

In the proposed algorithm, the prediction step consists in a first step to propagate at time $k$ the 
$\eta_{k-1}^N$ particles by simulating the transition kernel sampler, then in a second step to extend these 
particles according to a one-step-ahead sampler. 
Then the correction step consists in updating the weights of the $N$ particles
of $\eta_{k^-}^N$ according to their likelihood with  $y_k$ 
and the likelihood of their one-step ahead offspring particles with $y_{k+1}$.

The iteration $k-1\to k$ of the filter is more precisely:
\begin{description}

\item[\it Prediction step.]
As for the BPF we sample $N$ particles and compute $N$ normalized likelihood weights:
\begin{align*}
   \tilde{\xi}^{i}_k  \sim q_k(\,\cdot\,|\xi^{i}_{k-1}) \,, 
   \ 
   \tilde\omega^i_k   \propto \psi_k(\xi_{k^-}^i)\,\omega_{k^-}^i\,,
   \ 
   i=1:N\,.
\end{align*}
Again we will resample the particles $\tilde{\xi}^{1:N}_k$, but instead of doing so according to the weights $\tilde\omega^{1:N}_k$, we will first modify the latter ones. For each index $i$, we generate a one-step-ahead offspring particles and compute its weight:
\begin{align*}
  \tilde{\xi}^{i}_{k+1} 
     &\sim \tilde q_{k+1}(\,\cdot\,|\tilde{\xi}^{i}_k)\,,
  &
  \tilde{\omega}^{i}_{k+1} 
     &\eqdef \psi_{k+1}(\Tilde{\xi}^{i}_{k+1}) .
\end{align*}
Note that the likelihood weights $\tilde{\omega}^{i}_{k+1}$ depend on the next observation~$y_{k+1}$.
The choice of the one-step-ahead sampler $\tilde q_{k+1}$ will be detailed at the end of this sub-section.
\smallskip
\item[\it Correction step.]
We compute the weights at time $k$ according to Eq. \eqref{eq.smooth.corr}:
\begin{align*}
  \omega^i_{k} 
  &\eqdef 
  \tilde{\omega}^{i}_k\;\tilde{\omega}^{i}_{k+1}
   \qquad\textrm{for } i=1:N.
\end{align*}
Then,  $N$ particles  $\tilde{\xi}^{1:N}_k$ are resampled according to the weights $\omega^{1:N}_{k}$.

\end{description}
The PBPS is depicted in Algorithm \ref{algo.pbps}.

\paragraph{Choice of one-step-ahead sampler $\tilde q_{k+1}$}
For the special case of the system \eqref{eq.state.space.X}-\eqref{eq.state.space.Y} we can choose a simpler ``deterministic sampler'':
\[
 \tilde{\xi}^{i}_{k+1} = f_{k}(\tilde{\xi}^{i}_k)
\]
which corresponds to the one-step-ahead mean:
\begin{align*}
  \tilde{\xi}^{i}_{k+1}  
  &= \E\bigl(X_{k+1}\big|X_{k}=\tilde{\xi}^{i}_k\bigr)
\\
  &= \E\bigl(
         f_{k}(X_{k})+g_{k}(X_{k})\,W_{k}  
     \big|X_{k}=\tilde{\xi}^{i}_k\bigr)
\\
  &= f_{k}(\tilde{\xi}^{i}_k)+g_{k}(\tilde{\xi}^{i}_k)\,
     \underbrace{\E(W_{k}|X_{k}=\tilde{\xi}^{i}_k)}_{=0}  
  = f_{k}(\tilde{\xi}^{i}_k)\,.
\end{align*}
This choice greatly reduces the computation burden and it is sufficient, as we will see, for a linear state equation. For a highly nonlinear state equation, we can choose $\tilde q_{k+1}=q_{k+1}$, which corresponds 
to a simulation of the state dynamic.

\begin{listing}[t]
\small
\setstretch{1.3}
\begin{algorithmic}[1] 	    	    
  \STATE  $\xi^{1:N}_{0} \simiid \rho_{0}$ \COMMENT{\scriptsize initialization}
  \FOR {$k = 1,2,\dots$} 
    \FOR {$i = 1:N $}
       \STATE $\tilde{\xi}^{i}_k \sim q_k(\,\cdot\,|\xi^{i}_{k-1})$
            	\COMMENT{\scriptsize particles propagation}
       \STATE $\tilde{\omega}^{i}_{k}  
               \leftarrow \psi_{k}(\Tilde{\xi}^{i}_k)$      
	   \STATE $\tilde{\xi}^{i}_{k+1} 
	           = f_{k}(\Tilde{\xi}^{i}_k)$ 
            	\COMMENT{\scriptsize offspring } 
	   \STATE $\tilde{\omega}^{i}_{k+1} 
	           \leftarrow  \psi_{k+1}(\Tilde{\xi}^{i}_{k+1})$     
	           \COMMENT{\scriptsize offspring  weighting}
	   \STATE $\omega^i_{k} 
	           \leftarrow \tilde{\omega}^{i}_{k} \, \tilde{\omega}^{i}_{k+1}$\COMMENT{\scriptsize particles weighting}
  \ENDFOR
  \STATE $\omega^i_k 
          \leftarrow {\omega^i_k}/{\sum_{i'=1}^{N}\omega^{i'}_k} \,,\  i=1:N$ 
          \COMMENT{\scriptsize weights normalization}  		
  \STATE $\xi_k^{1:N} \simiid \sum_{i'=1}^{N}
          \omega^{i'}_k\,\delta_{\Tilde{\xi}^{i'}_k}$ 
          \COMMENT{\scriptsize particles resampling}	
  \RETURN $\xi_k^{1:N}$ 
\ENDFOR
\end{algorithmic}
\caption{predictive bootstrap particle smoother (PBPS).}
\label{algo.pbps}
\end{listing}

\subsection{The regime switching predictive bootstrap particle smoother}
\label{sec.robust}

\begin{listing}
\small
	\setstretch{1.3}
	\begin{algorithmic}[1] 	    	    
		\STATE $\xi^{1:N}_{0} \simiid \rho_{0}$ 
            	 \COMMENT{\scriptsize initialization}
		\RETURN $\xi_0^{1:N}$
		\FOR{$k = 1,2,\dots$}  
		\STATE $\mu^{1:N}_{k-1} \simiid U(\MM)$
		\STATE $\tilde{\xi}^{i}_k \sim q_k(\,\cdot\,|\xi^{i}_{k-1},\mu^{i}_{k-1}) \,,\  i=1:N $ 
		\COMMENT{\scriptsize particles propagation}
		\STATE $\tilde{\omega}^{i}_{k}  
		\leftarrow \psi_{k}(\Tilde{\xi}^{i}_k) \,,\  i=1:N $       
		\STATE $\tilde{\xi}^{i}_{k+1} 
		= f_{k}(\Tilde{\xi}^{i}_k)\,,\  i=1:N $ 
		\COMMENT{\scriptsize offspring}
		\STATE $\tilde{\omega}^{i}_{k+1} 
		\leftarrow  \psi_{k+1}(\Tilde{\xi}^{i}_{k+1})\,,\  i=1:N $    
		\COMMENT{\scriptsize offspring weighting}
		\STATE $\omega^i_{k} 
		\leftarrow \tilde{\omega}^{i}_{k} \, \tilde{\omega}^{i}_{k+1}\,,\  i=1:N $
		\COMMENT{\scriptsize particles weighting}

		\STATE $\omega^i_k \leftarrow  {\omega^i_k}/{\sum_{j=1}^{N}\omega^{j}_k} \,,\  i=1:N$ 
		\COMMENT{\scriptsize renormalization}
		\STATE $\xi_k^{1:N} \simiid \sum_{j=1}^{N}\omega^{j}_k\,\delta_{\tilde{\xi}^{j}_{k}}$
		\COMMENT{\scriptsize resampling} 	
		\RETURN $\xi_k^{1:N}$	 
		\ENDFOR
	\end{algorithmic}
\caption{Regime Switching predictive bootstrap particle smoother  (RS-PBPS)}
\label{algo.rs-pbps}
\end{listing}

In most applications, both the state model and the observation model are only partially known. As we will see in Section \ref{sec.ex2}, applying a filter without care in such situations can be delicate. There are several techniques to make these filters more robust to model mismatch such as the dynamic model averaging  (DMA) \cite{liu2019,urteaga2016}  or the regime switching (RS) \cite{ellaham2021}.
We will use both DMA  and RS techniques in the simulations of Section \ref{sec.ex2}, 
but we introduce only the RS technique in the present section by
proposing a ``robustified'' version of PBPS, called regime switching PBPS (RS-PBPS).

\medskip

The partial knowledge on the state-space model introduced in 
Section \ref{sec.state.space.model} 
can be related to the state model, i.e. to the the transition kernel $q_{k}$, or  the observation model, i.e. the likelihood function $\psi_{k}$.

In the great majority of applications, the uncertainty relates to the state model, so we will restrict ourselves to this case.
In order to model this uncertainty, we assume that the state dynamics corresponds to a transition kernel of the form:
\[
   q_{k}(x|x',m)\,,
\]
where $m$ is a parameter that evolves dynamically in the finite set:
\[
 \mathcal{M}=\{m_{1},\dots,m_{L}\}\,.
\]
Thus $\mathcal{M}$ represents the different possible regimes of the state model over time.

The principle of these robust filters consists in making the particles and the weights evolve according to a mixture of the parameters $\mathcal{M}$. 
The dynamic model averaging particle filter (DMA-BPF) proposed in \cite{urteaga2016}  is detailed in Algorithm \ref{algo.dma-bpf}, 
we will not describe it here.

We now present one of the algorithms proposed by \cite{ellaham2021}. We use the simplest of them which gives good results according 
to \cite{ellaham2021}. In the BPF, we replace the prediction step (Algorithm \ref{algo.bpf} line 4) by:
\[
   \mu^{1:N}_{k-1} \simiid U(\MM)
   \,,\quad
   \xi^i_k \sim q_k(\,\cdot\,|\xi^i_{k-1},\mu^i_{k-1})
   \,,\quad i=1:N\,,
\]
where $U(\MM)$ is the uniform distribution on $\MM$;
the only difference lies in the fact that  the particles are sampled 
at random according to the various models represented by $\MM$. The resulting method called RS-BPF is described in Algorithm \ref{algo.rs-bpf}.
The adaptation to PBPS and APF  is also immediate, leading to the RS-PBPS and the RS-APF; the RS-PBPS is detailed in Algorithm~\ref{algo.rs-pbps}.

\section{Simulation studies}
\label{sec.simulation}

We compare numerically the PBPS to other filters on two space-state models 
where observability is weakly conditioned. 
The performance of the filters is compared using an empirical evaluation of the root-mean-square error (RMSE). 
We simulate $S$ independent trajectories $(X_{0:K}^{(s)},Y_{1:K}^{(s)})_{s=1:S}$ of 
the state-space models, where $K$ is the length of the chronological sequence of observations. For each simulation $s$, we ran $R$ times each  filter $\FF\in\{$BPF, PBPS, APF, DMA-BPF, RS-BPF, RS-APF, RS-PBPS$\}$ and we compute the root mean squared error at time~$k$:
\begin{align}
	\label{eq:simulation:2} 
	\textrm{RMSE}_k(\FF)
	&
	\eqdef 
        \sqrt{	
        \frac{1}{S} 
        \sum_{s=1}^{S}
	       \frac{1}{R} \sum_{r=1}^{R} \Bigl|\hat{X}^{\FF(s,r)}_k - X^{(s)}_k\Bigr|^2}\,,
\end{align}
where:
\begin{align*}
	\hat{X}_k^{\FF(s,r)}
	&\eqdef  \int_{\mathbb{R}^{n}} x\,\eta_k^{N,\FF(s,r)}(x)\,\rmd x
	= 
	\frac{1}{N}\sum_{i=1}^{N}\xi_k^{\FF(s,r),i}
\end{align*}
is the numerical approximation of $\hat X^{(s)}_{k}=\E(X^{(s)}_{k}|Y^{(s)}_{1:k})$ by the filter $\FF$.
We also compute the global root mean squared error:  
\begin{align*}
	\textrm{RMSE}(\FF)
	&
	\eqdef 
	\frac{1}{K} \sum_{k=1}^{K}\textrm{RMSE}_k(\FF)
	\,.
\end{align*}
Note that for the EKF and UKF filters the  summation over $r$ in \eqref{eq:simulation:2}  is useless.

\bigskip

All implementations are done in R language \cite{r2021a} using a 2.11 GHz core i7-8650U intel running Windows 10 64bits with a 16 Go RAM.

\subsection{First case study: a one-dimen\-sional model}
\label{sec.ex1}

We consider the  following one-dimensional nonlinear  model  \cite{gordon1993a, kitagawa1996a,doucet2001a}:
\begin{align}
	\label{eq:simulation:3}
	\begin{split}
		X_{k} 
		&=
		\frac{X_{k-1}}{2} 
		+ \frac{25\,X_{k-1}}{1 + X_{k-1}^2} 
		+ 8\cos\bigl(1.2\,(k-1)\bigr) + W_{k-1}\,,
	\\
		Y_{k} 
		&= \frac{X_{k}^2}{20} + V_{k}\,,
	\end{split}
\end{align}
with $ 1\leq k \leq K$ ($K=50$),  $X_{0} \sim \NN(0, 1)$; $W_k \simiid \NN(0, 3^2)$, $V_k \simiid  \NN(0, 1)$; $X_{0}$, $(W_{k})_{k\geq 1}$ and $(V_{k})_{k\geq 1}$ mutually independent. Note that the state process $X_{k}$ is observed only through $X_k^2$ so the filters have difficulties in determining whether $X_{k}$ is 
positive or negative, especially since the state process $X_{k}$ 
regularly changes it's sign. Thus this model is regularly used 
as benchmark for testing filters, as filters may easily lose track of $X_{k}$.

For this example we compare the filters PBPS, BPF, APF, EKF and UKF. 

\bigskip

\begin{figure}
\begin{center}
\includegraphics[width=\columnwidth]{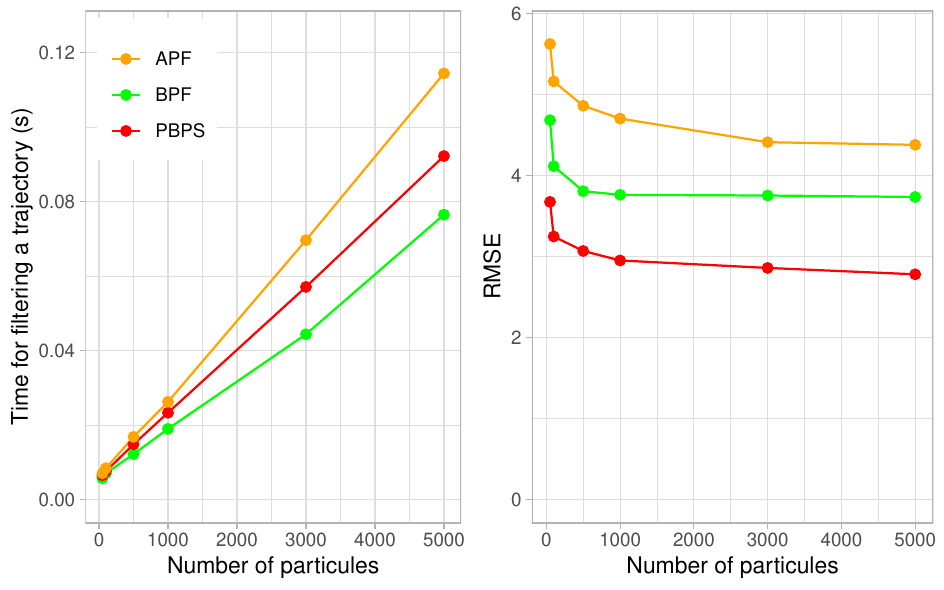}
\end{center}
\caption{
One-dimensional case study \eqref{eq:simulation:3} ---
We plot the average computation time for filtering a trajectory (left) and the RMSE (right) 
of the filters  PBPS,  BPF and  APF  as a function of different values of $N$ 
(50, 100, 500, 1000, 3000, 5000) (with $S=100$ and $R = 40$).
In comparison, the EKF and UKF have a respective RMSE of 16.83 and 6.88 
(and a negligible computation time compared to the particle approximations).
}

\label{fig:simulationMod1:0}
\end{figure}

In Figure \ref{fig:simulationMod1:0} we plot, as a function of different values of $N$ (50, 100, 500, 1000, 3000, 5000), on the left the average computation time for the filtering a trajectory, and on the right the RMSE (with $S=100$, $R = 40$). 

The computation times of EKF and UKF are negligible compared to those of the particle approximations. On the other hand, the RMSE of EKF (16.83) are much higher than those of the particle approximations; UKF with an RMSE of 6.88 behaves much better than EKF but is still less accurate than the particle approximations.

The computation time of PBPS is naturally higher than that of BPF, but the latter is much less accurate. For example, PBPS with $N=50$ particles is 15 times faster than BPF with $N=5000$ particles while being 30\%\ more accurate.

\begin{figure*}[h]
\begin{center}
\includegraphics[scale=0.8]{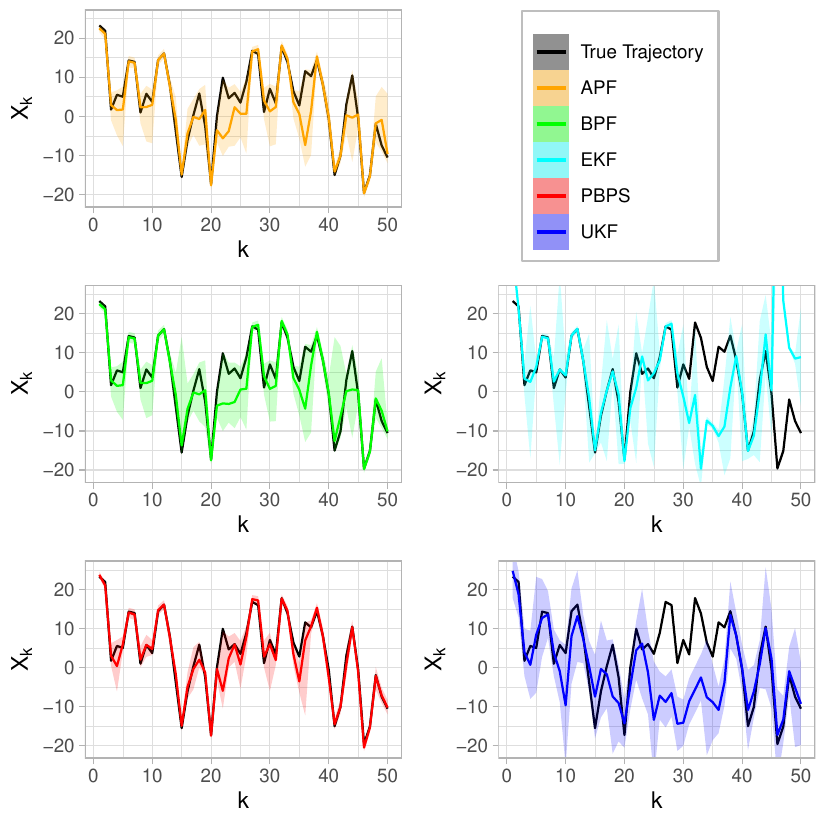}
\end{center}
\caption{
One-dimensional case study \eqref{eq:simulation:3} --- On a single simulation, 
for each filter, we plot the true state trajectory $k\to X_{k}$ and the approximation $k\to \hat{X}_k$ 
with the associated $95\%$ confidence region ($N=5000$).
}
\label{fig:simulation:2}
\end{figure*}

\bigskip

In Figure~\ref{fig:simulation:2}, for each filter 
$\FF\in\{\textrm{PBPS, BPF, APF, EKF, UKF}\}$ (with $N=5000$), we plot the true state trajectory $k\to X_{k}$, the approximation $k\to \hat{X}^{\FF}_k$, and the associated 95\%\ confidence regions.  For the particle approximations the confidence interval is given empirically by the particles, for the EKF and UKF it is given as an approximated Gaussian confidence interval.
The PBPS approximation has a smaller error $k\to |\hat{X}^{\FF}_k-X_{k}|$ and a smaller confidence region than the other filters.

\bigskip

\begin{figure}
\begin{center}
\includegraphics[scale=0.8]{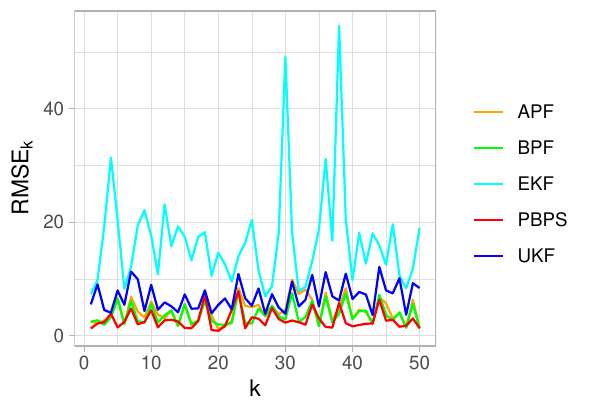}
\end{center}
\caption{One-dimensional case study \eqref{eq:simulation:3} ---
We compare $k\to \textrm{RMSE}_k$ for each filter ($S=100$, $R=40$, $N=5000$).}
\label{fig.ex1.rmse.k}
\end{figure}

In Figure \ref{fig.ex1.rmse.k}, we  compare $k\to \textrm{RMSE}_{k}(\FF)$ 
for $\FF\in\{$PBPS, BPF, APF, EKF, UKF$\}$ (with $N=5000$). 
Again, we see that the PBPS behaves better than the other filters and that the 
EKF has a very erratic behavior.

\bigskip

The poor behavior of the EKF can be explained by the very nonlinear nature of this example and by the observability problem around 0. Note the relatively good behavior of the UKF. However, because of the observability problem at 0, both the UKF and the EKF cannot find the track of $X_k$ once in the wrong half-plane (Figure \ref{fig:simulation:2} around time $k=30$). 

\begin{figure}[h]
\begin{center}
\includegraphics[scale=0.7]{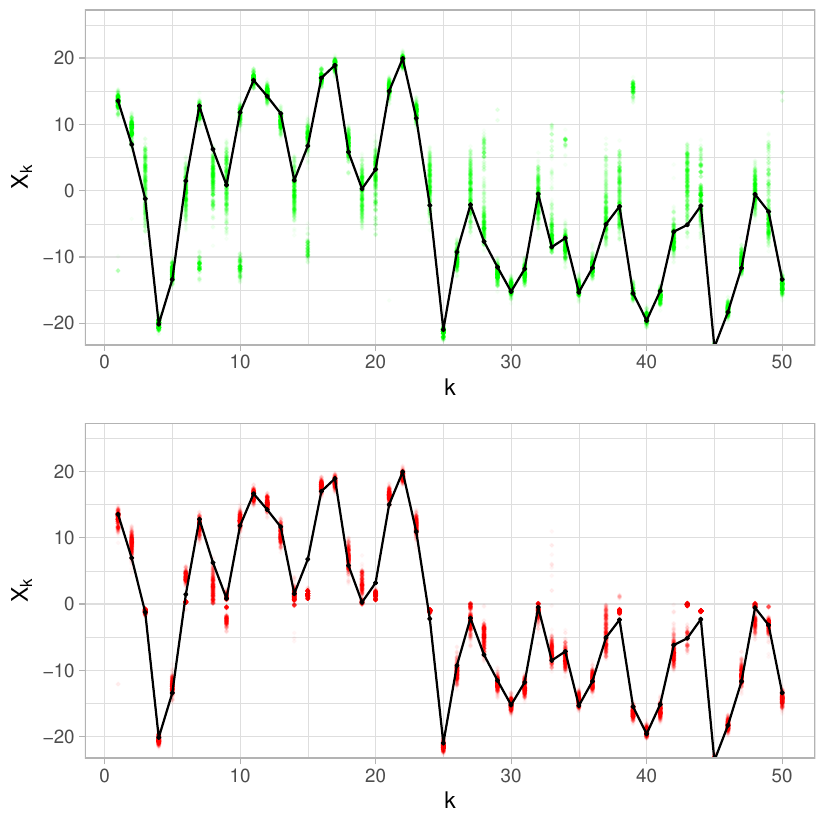}
\end{center}
\caption{One-dimensional case study \eqref{eq:simulation:3} ---
On a single simulation, we plot the true value $k\to X_k$ with the evolution of the set of $N=500$ particles $k\to\xi^{1:N}_k$ for the BPF (top) and the PBPS (bottom); 
the particles are represented in transparency to better reflect their density.}
\label{fig.ex1.particles}
\end{figure}

In Figure \ref{fig.ex1.particles}, 
in the case of a single simulation of the state-space system and of the BPF and PBPS filters (with $N=5000$), 
we plot the true trajectory $k\to X_k$ as well as the set of particles $k\to \xi_k^{\FF,1:N}$ 
for $\FF\in\{$BPF, PBPS$\}$. Not surprisingly, we find that the PBPS particles are more 
concentrated around the true value $X_k$ and that BPF more often ``loads'' the symmetric part 
of the state space. Indeed, the smoothing allows us to reduce the variance but also to compensate  
the observability ambiguities; it is for this purpose that it was designed, and we reiterate that  
this gain is obtained with a great improvement in the computation time (provided 
that less particles are used).

\subsection{Second case study: a four-dimensional bearings-only tracking model}
\label{sec.ex2}

We consider the  following four-dimensional model \cite{gordon1993a,pitt1999a,campillo2009a}: 
\begin{align}
	\label{eq:simulation:5}
	\begin{split}
X_k 
&= \left(\begin{smallmatrix} 
	1 & 0 & 1 & 0 \\
	0 & 1 & 0 & 1 \\
	0 & 0 & 1 & 0 \\
	0 & 0 & 0 & 1 
\end{smallmatrix}\right)\,X_{k-1} 
+ 
\sigma_{W}\,\left(\begin{smallmatrix} 
	0.5 & 0 \\
	0 & 0.5 \\
	1 &  0\\
	0 & 1 \\
\end{smallmatrix}\right)\,W_{k-1}\,,
\\
Y_{k} 
&\sim \textrm{wrapped Cauchy}\Bigl(\arctan\Bigl(\frac{X_k[1]}{X_k[2]}\Bigr),\rho\Bigr),
	\end{split}
\end{align}
with $1\leq k\leq K=20$, $X_0 \sim \NN(\bar X_{0},P_0)$ with:
\begin{align*}
 &\bar X_0= \bigl(-0.05,  0.2,0.001, -0.055\bigr)^*\,,
 \\
 &P_0= 0.01\, \textrm{diag} \bigl(0.5^2,  0.3^2, 0.005^2,0.01^2\bigr)\,,
\end{align*}
$\sigma_{W} =0.001$, $W_k\simiid \NN(0,I_{2\times 2})$,  $\rho=1 - 0.005^2$,  $X_{0}$ and $(W_k)_{k\geq 1}$ mutually independent; conditionally on $(X_k)_{k\geq 0}$, $(Y_k)_{k\geq 1}$ and $(W_k)_{k\geq 1}$ are  
independent.

The state equation corresponds to a target moving on a plane. The state vector is:
\[
  X_k 
  = 
  (X_k[1],X_k[2],X_k[3],X_k[4])^* 
  = (x_{1},x_{2}, \dot x_{1},\dot x_{2})^*\,,
\]
where $(x_{1},x_{2})$ are the Cartesian coordinates of the target in the plane and $(\dot x_{1},\dot x_{2})$  are the corresponding velocities. 
The observer is located at the origin of the plane and accessed only to the azimuth angle $\beta=\arctan(x_1/x_2)\in[-\pi,\pi]$ corrupted by noise. 

\medskip

It is known that in the situation where the observer follows a rectilinear trajectory at constant speed, or 
zero speed as here, the filtering problem is poorly conditioned \cite{nardone1981a}. This is a situation 
where it is known for example 
that the EKF diverges very strongly. 

\medskip

The conditional probability density function of the 
measured angle $Y_{k}$ given the state $X_{k}$ is assumed to be a wrapped Cauchy distribution with concentration parameter $\rho$~\cite{pitt1999a}:	
\begin{equation}
\label{eq:simulation:5.Y}
   p_{Y_k|X_k=x}(y)
   =
   \frac
   {(2\,\pi)^{-1}\,(1-\rho^2)}
   {
     1
     +
     \rho^2-2\,\rho\,\cos\Bigl(y-\arctan(\frac{x[1]}{x[2]}))\Bigr)
     }
\end{equation}
for $-\pi\leq y<\pi$,
where  $\rho \in [0,1]$ is the mean resultant length. Note that the state dynamics is linear and Gaussian, but the observation 
dynamics is nonlinear and non-Gaussian.

\bigskip

Classically, in tracking studies, the trajectory model (the state equation) does not correspond to reality. 
In reality, the target follows a uniform straight trajectory sometimes interrupted by changes in heading and/or speed. The variance $\sigma_W$ of the state equation  \eqref{eq:simulation:5} appears then as a parameter of the filter: if it is too small, the filter will have trouble tracking the target, if it is too large, the particles will spread out too much, the filter will then lose accuracy, and even lose  track of the target. 

For bearings-only  tracking applications, the EKF and UKF filters have poor performances 
and will therefore not be considered in this example. On the other hand, the robust filters 
introduced in Section \ref{sec.robust} are relevant here because 
there is a mismatch model.

We compare the filters in two scenarios (with $K=40$ time steps):
\begin{itemize}
\item the target follows a uniform rectilinear motion;
\item the target follows a uniform rectilinear motion with a change of heading at the middle of the simulation.
\end{itemize}
As we will see, the first scenario is simpler than the second.
For the filter we consider two different values for $\sigma_W$: a  small value $\sigma_{W}=0.001$ and a large $\sigma_{W}=0.003$. $\sigma_{W}=0.001$ is in fact too small to be consistent with the simulated trajectory, 
the second value $\sigma_{W}=0.003$ is more consistent with the simulated trajectory.
For the robust filter we consider:
\[
   \MM = \{0.0005,\,0.001,\,0.003,\,0.005\}
\]
as the set of possible models. 

\bigskip

\begin{figure}
\begin{center}
\includegraphics[scale=0.8]{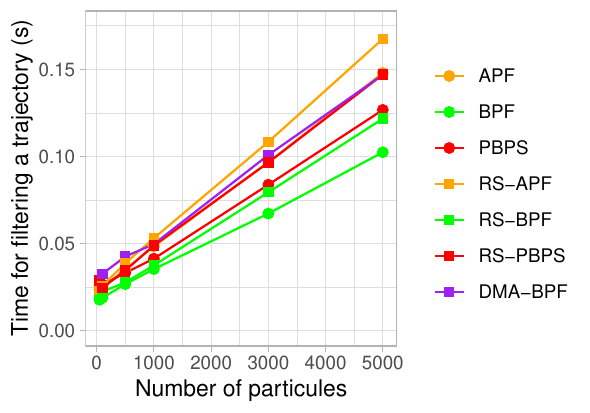}
\end{center}
\caption{
Four-dimensional case study \eqref{eq:simulation:5.Y} ---
We plot the average computation time for filtering a trajectory
of the filters  PBPS,  BPF, APF,  RS-PBPS,  RS-BPF, RS-APF and  DMA-BPF  as a function of different values of $N$ 
(50, 100, 500, 1000, 3000, 5000) (with $S=20$ and $R = 50$).
}
\label{fig.ex2.temps}
\end{figure}

\begin{figure*}
\begin{center}
\includegraphics[scale=0.8]{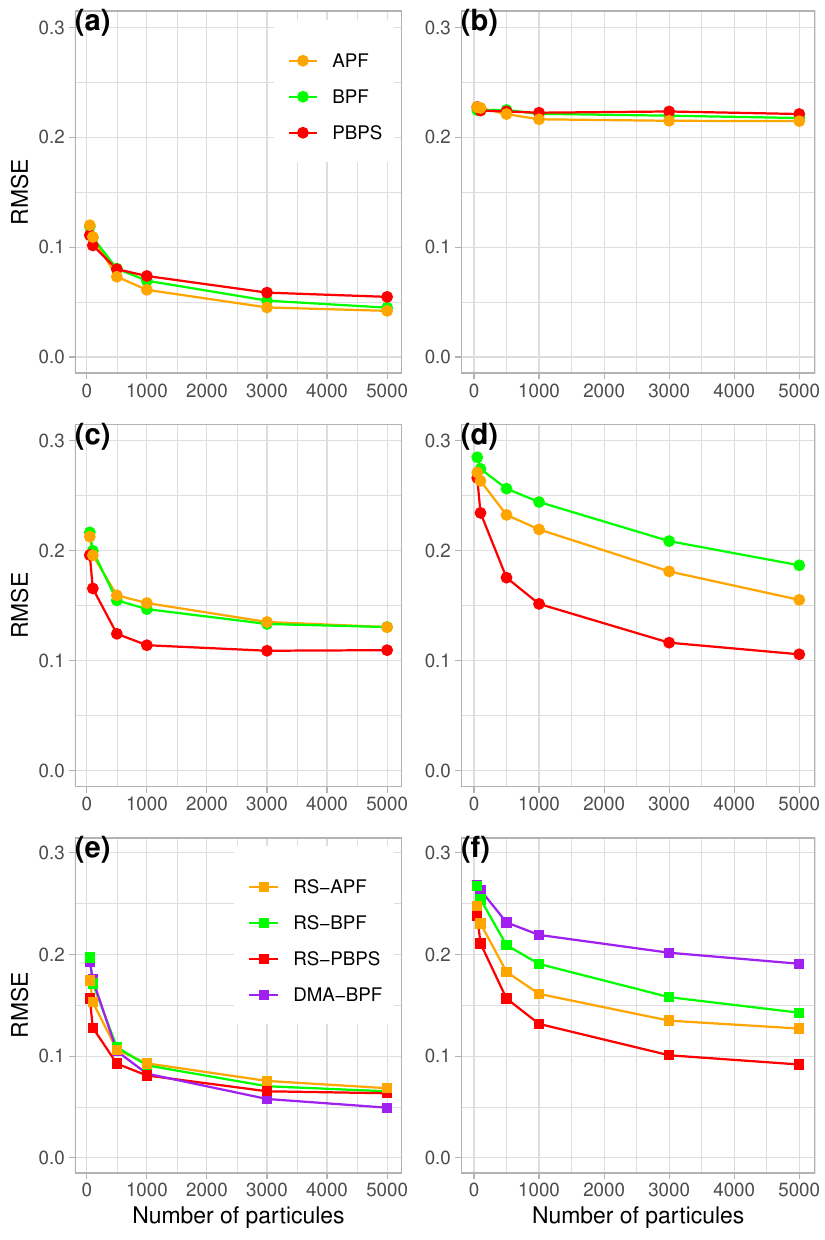}
\end{center}
\caption{
Four-dimensional case study \eqref{eq:simulation:5.Y} ---
Evolution 
of the RMSE as a function of the number $N$ of particles: the left column corresponds to the 
case without a turn, the right column to the case with a turn. 
For PBPS, BPF and APF, the first row (top) presents the RMSE for with $\sigma_{W}=0.001$, 
the second (middle) with $\sigma_{W}=0.003$. The third row (bottom) shows the RMSE for 
the RS-PBPS, RS-BPF, RS-APF, and DMA-BPF filters.}
\label{fig.ex2.rmse}
\end{figure*}

\begin{figure*}
\begin{center}
\includegraphics[scale=0.8]{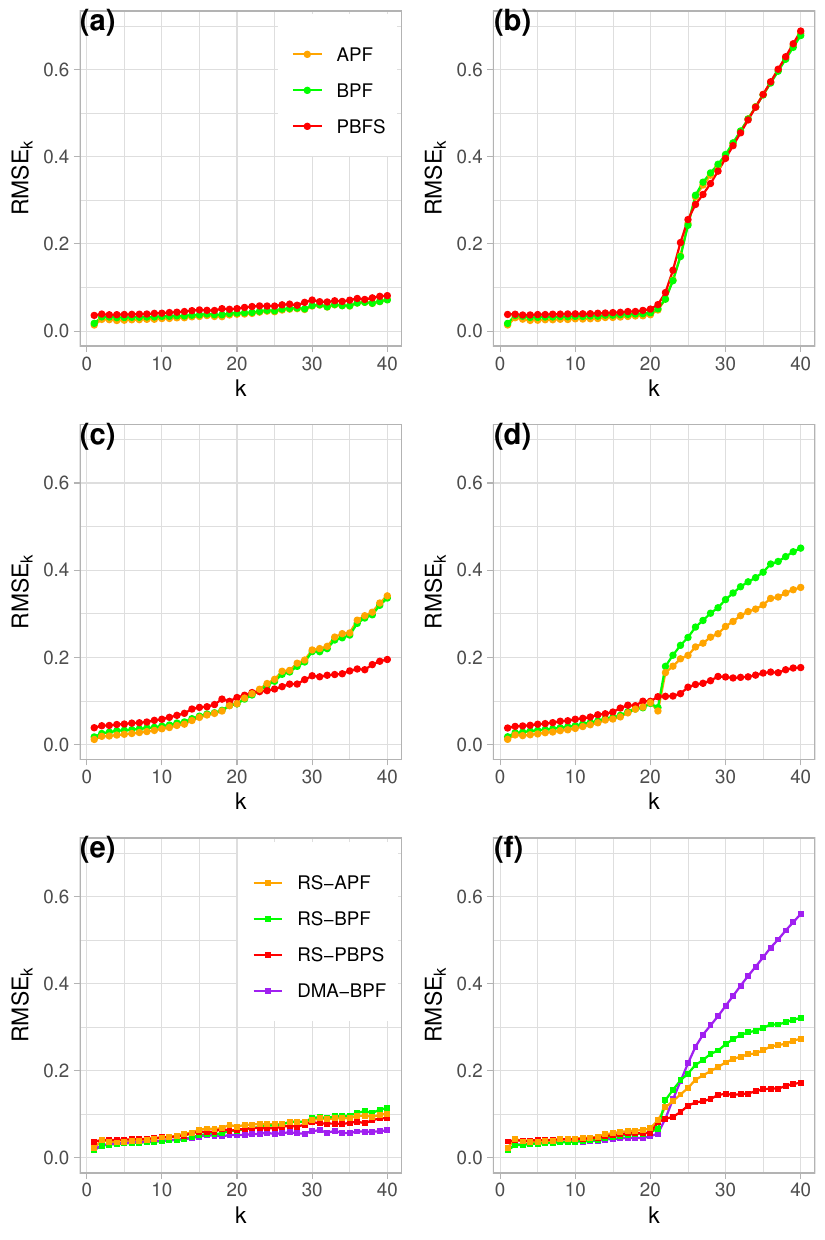}
\end{center}
\caption{
Four-dimensional case study \eqref{eq:simulation:5.Y} ---
$k\to $RMSE$_k$ (for $N=5000$, $R=50$, $S=20$)
for the different filters:
the left column corresponds to the 
case without a turn, the right column to the case with a turn; 
PBPS, BPF and APF for with $\sigma_{W}=0.001$, on the first row;
PBPS, BPF and APF for with $\sigma_{W}=0.003$, on the middle row;
RS-PBPS, RS-BPF, RS-APF, and DMA-BPF, on the bottom row.
}
\label{fig.ex2.rmse.k}
\end{figure*}

\begin{figure*}
\begin{center}
\includegraphics[scale=0.8]{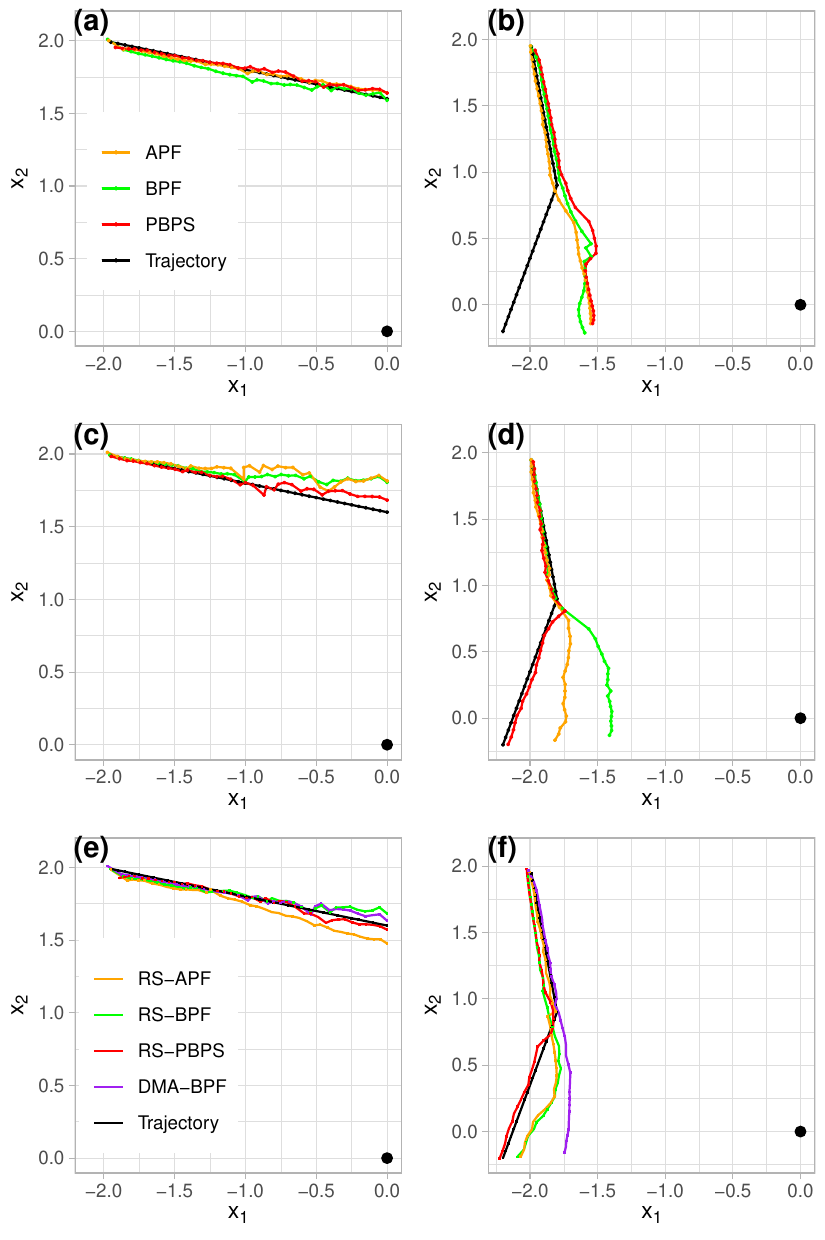}
\end{center}
\caption{
Four-dimensional case study \eqref{eq:simulation:5.Y} ---
On a single simulation and for each filter, we plot the true trajectory of the target 
with the estimates of each filter: without turn in the left column, with 1 turn in the right column; 
PBPS, BPS and APF filters with small $\sigma_W=0.001$ in the top row,
PBPS, BPS and APF filters with large $\sigma_W=0.003$ in the middle row,
RS-PBPS, RS-BPS, RS-APF, and DMA-BPF filters in the bottom row. The observer is in position $(0,0)$.}
\label{fig.ex2.estimates}
\end{figure*}

In Figure \ref{fig.ex2.temps} we present the average computation times for 
filtering a trajectory as a function of different values of $N$ 
(50, 100, 500, 1000, 3000, 5000) (with $S=20$ and $R = 50$).

In Figure \ref{fig.ex2.rmse}, in agreement with Figure \ref{fig.ex2.temps}, we present the evolution 
of the RMSE as a function of the number $N$ of particles: the left column corresponds to the 
case without a turn, the right column to the case with a turn. 
For PBPS, BPF and APF, the first row (top) presents the RMSE for with $\sigma_{W}=0.001$, 
the second (middle) with $\sigma_{W}=0.003$. The third row (bottom) shows the RMSE for 
the RS-PBPS, RS-BPF, RS-APF, and DMA-BPF filters.

In Figure \ref{fig.ex2.rmse.k}  we plot $k\to\ $RMSE$_k$ (for $N=5000$, $R=50$, $S=20$)
for the different filters:
the left column corresponds to the 
case without a turn, the right column to the case with a turn; 
PBPS, BPF and APF for with $\sigma_{W}=0.001$, on the first row;
PBPS, BPF and APF for with $\sigma_{W}=0.003$, on the middle row;
RS-PBPS, RS-BPF, RS-APF, and DMA-BPF, on the bottom row.

In Figure \ref{fig.ex2.estimates}, on a single simulation and for each filter, we plot the true trajectory of the target 
with the estimates of each filter: without turn in the left column, with 1 turn in the right column;
PBPS, BPS and APF filters with small $\sigma_W=0.001$ in the top row,
PBPS, BPS and APF filters with large $\sigma_W=0.003$ in the middle row,
RS-PBPS, RS-BPS, RS-APF, and DMA-BPF filters in the bottom row.

\bigskip

According to Figure \ref{fig.ex2.temps}, at the same number of particles, 
BPF is slightly faster than PBPS, both being faster than APF. The 
overheads for the robust version (RS) are also reasonable. In terms 
of complexity, the computation time is linear with the number of particles.

In order to better understand the situation, we will first comment on Figures~\ref{fig.ex2.rmse.k} and \ref{fig.ex2.estimates}: 
with a too small $\sigma_W$, the filters are able to track the target when it stays in a straight line (a) but they are 
not able to adapt when a turn occurs and continue the tracking more or less in a straight line (b). With a larger 
$\sigma_W$, the filters are still able to track the target
 when it remains in a straight line (c) 
with a small and increasing inaccuracy; in contrast only PBPS behaves reasonably well when a turn occurs (d). Except DMA-BPF which has a bad behavior, the RS robust filters 
behave better and RS-PBPS remains better than all the others (e) and (f).

From Figures \ref{fig.ex2.temps} and \ref{fig.ex2.rmse}, in the case of a turn, which is the most complex situation, we see that in order to reach PBPS/RS-PBPS accuracy 
with $N=1000$ the other filters must use $N=5000$ particles, 
and even in this case PBPS/RS-PBPS is at least twice as fast as the others.

\section{Discussion and conclusion}

In this paper, our aim was to propose a particle approximation method, not much more complex than the boostrap particle filter, which reduces the error variance, while being more robust in weakly conditioned situations, especially when observability conditions are poor.

We have therefore proposed an new particle filter algorithm, called predictive bootstrap particle smoother (PBPS), that 
deviates slightly from the strict filtering method in the sense that 
it takes into account the next observation in addition to the current one. 
We can thus consider PBPS as an approximation of the smoother with a 
fixed-lag interval of one step. However, the proposed algorithm is considerably simpler than the smoothing algorithms.

Furthermore, a way of making a filter more robust is to extend it to a ``regime switching'' strategy framework as proposed in \cite{ellaham2021}; which is what we have done by proposing a robustified version of PBPS.

In the considered examples, EKF shows very poor performance, which is 
perfectly understandable due to the very nonlinear nature and the poor 
observability conditions of the examples. It should be noted that UKF 
performs much better than EKF while remaining far from the performance 
of particle filters. Among the particle filters, APF performs slightly less well than
 than PBF and PBPS. Finally, even if for a given number of particles, 
BPF is faster than PBPS, the latter presents both a lower error 
variance and a much better behavior regarding observability problems. 
In terms of accuracy, PBPS with 1000 particles is still better than BPF 
with 5000 particles while being much faster. This is also the case when 
comparing RS-PBPS with RS-BPF

These performances are due to the fact that PBPS is a very simplified 
approximation of the one time step fixed-lag smoother. A strategy using 
a deeper fixed-lag interval of 2 or more time steps would be too costly 
and would not allow us to recover such performances.

In view of the performance of PBPS it would be interesting to develop 
strategies where the number $N$ of particles adapts dynamically to the 
problem conditions. It would also be interesting to test the PBPS 
strategy in other cases, such as when the time between two consecutive 
observations is important and requires several successive prediction steps.

\appendix\footnotesize
\def\thesection{A}

\section*{Appendix: Algorithms}

\begin{listing}[H]
	\setstretch{1.3}
	\begin{algorithmic}[1] 	    	    
		\STATE $\xi^{1:N}_{0} \simiid \rho_{0}$ 
		\COMMENT{\scriptsize initialization}
		\RETURN $\xi_0^{1:N}$
		\FOR{$k = 1:K$}  
		\STATE $\bar\xi^{i}_{k} = \E(X_{k}|X_{k-1}=\xi^{i}_{k-1})  \,,\  i=1:N $  
		\COMMENT{\scriptsize mean particle evolution}
		\STATE $\bar\omega_k^i \leftarrow \psi_{k}(\bar\xi^{i}_{k}) \,,\  i=1:N $ 
		\COMMENT{\scriptsize  ... and weighting}
		\STATE $\tilde\xi_{k-1}^{1:N} \simiid \sum_{j=1}^{N}\bar\omega_k^i \,
		                          \delta_{\xi^{j}_{k-1}}$
		\COMMENT{\scriptsize initial particles resampling}
		\STATE $\xi^{i}_{k^-} \sim q_k(\,\cdot\,|\tilde\xi^{i}_{k-1})  \,,\  i=1:N $ 
		\COMMENT{\scriptsize particles propagation}
		\STATE $\omega^i_{k} \leftarrow 
		        \psi_{k}(\xi^{i}_{k^-})/\bar\omega_k^i \,,\  i=1:N $ 
		\COMMENT{\scriptsize likelihood}
		\STATE $\omega^i_k \leftarrow  {\omega^i_k}/{\sum_{j=1}^{N}\omega^{j}_k} \,,\  i=1:N$ 
		\COMMENT{\scriptsize renormalization}
		\STATE $\xi_k^{1:N} \simiid \sum_{j=1}^{N}\omega^{j}_k\,\delta_{\xi^{j}_{k^-}}$
		\COMMENT{\scriptsize resampling} 	
		\RETURN $\xi_k^{1:N}$	 
		\ENDFOR
	\end{algorithmic}
	\caption{Auxiliary particle filter (APF).
	Note that in the case of the \eqref{eq.state.space.X}-\eqref{eq.state.space.Y}, 
	the first step, line 4, is reduced to 
	$\bar\xi^{i}_{k} = \E(X_{k}|X_{k-1}=\xi^{i}_{k-1})=f_{k-1}(\xi^{i}_{k-1})$}
	\label{algo.apf}
\end{listing}

\begin{listing}[H]
\setstretch{1.3}
\begin{algorithmic}[1] 	    	    
	\STATE $\xi^{1:N}_{0} \simiid \rho_{0}$,
	$\mu^{1:N}_{0} \simiid U(\MM)$ 
	\COMMENT{\scriptsize initialization}
	\RETURN $\xi_0^{1:N}$
	\FOR{$k = 1:K$}  
		\STATE 
		$\xi^{i}_{k^*} \sim q_k(\,\cdot\,|\xi^{i}_{k-1},\mu^{i}_{k-1})$, $i=1:N $
		\STATE
		$\omega^{l}_k=\sum_{i=1}^{N} \psi_{k}(\xi^{i}_{k^*}) 
		    \indic_{\mu^{i}_{k-1}=m_l}$,
		    $l=1:L $
		     \COMMENT{\scriptsize likelihood of candidate models}
		\STATE
		$N_{k}^{1:L} \simiid 
		   \textrm{Multinomial}(N;m_{1},\dots,m_{L};\omega^{1}_k,\cdots,\omega^{L}_k)$\\ 
		with $N_{k}^{1}+\dots+N_{k}^{L}=N$
		  \COMMENT{\scriptsize number of particles per candidate model}
		\STATE
		$\hat\xi^{1,1:L}_{k-1} 
	 		\simiid  
			\sum_{j=1}^{N} \delta_{\xi^{j}_{k-1}}\,
			\delta_{\mu^{j}_{k-1}=m_{l}} $
		 \COMMENT{\scriptsize initial particles resampling}

		 \vskip-0.5em
		 \hphantom{dgg}
		 \COMMENT{\scriptsize per candidate model}
		\STATE 
		$(\tilde\xi_{k-1}^{1:N},\mu_k^{1:N})
		   \leftarrow(\hat\xi^{1:N_{k}^{1},1}_{k-1},m_{1}),(\hat\xi^{1:N_{k}^{2},2}_{k-1},m_{2}),\dots,(\hat\xi^{1:N_{k}^{L},L}_{k-1},m_{L})$ 
		\STATE $\xi^{i}_{k} \sim q_k(\,\cdot\,|\tilde\xi^{i}_{k-1},\mu^{i}_{k})  \,,\  i=1:N $ 
		\COMMENT{\scriptsize particles propagation}
		\RETURN $\xi_k^{1:N}$	 
	\ENDFOR
\end{algorithmic}
\caption{Dynamic model averaging bootstrap particle filter (DMA-BPF) \cite{urteaga2016}.}
\label{algo.dma-bpf}
\end{listing}

\begin{listing}[H]
	\setstretch{1.3}
		\begin{algorithmic}[1] 	    	    
		\STATE $\xi^{1:N}_{0} \simiid \rho_{0}$ 	
			\COMMENT{\scriptsize initialization}
		\RETURN $\xi_0^{1:N}$
		\FOR{$k = 1:K$}  
		    \STATE $\mu^{1:N}_{k-1} \simiid U(\MM)$
		    \STATE $\xi^{i}_{k^-} \sim q_k(\,\cdot\,|\xi^{i}_{k-1},\mu^{i}_{k-1})$, $i=1:N $ 	
		\COMMENT{\scriptsize particles propagation}
		    \STATE $\omega^i_{k} \leftarrow \psi_{k}(\xi^{i}_{k^-}) \,,\  i=1:N $ 
		\COMMENT{\scriptsize likelihood}
		    \STATE $\omega^i_k \leftarrow  {\omega^i_k}/{\sum_{j=1}^{N}\omega^{j}_k}$, $i=1:N$ 
		\COMMENT{\scriptsize renormalization}
		    \STATE $\xi_k^{1:N} \simiid \sum_{j=1}^{N}\omega^{j}_k\,\delta_{\xi^{j}_{k^-}}$
		\COMMENT{\scriptsize resampling} 	
		    \RETURN $\xi_k^{1:N}$	 
		\ENDFOR
	\end{algorithmic}
	\caption{Regime Switching Bootstrap Particle Filter (RS-BPF)~\cite{ellaham2021}.}
	\label{algo.rs-bpf}
\end{listing}




\printbibheading
\printbibliography[heading=subbibliography,nottype=software,nottype=softwareversion,nottype=softwaremodule,nottype=codefragment,heading=none]

\end{document}